\documentclass{article}
\pdfoutput=1 

\usepackage{jinstpub} 

\title{\rightline{\rm\normalsize IIT-CAPP-17-4}{\boldmath Improved performance of semiconductor laser tracking frequency gauge}}

\author[a,1]{D.M. Kaplan,\note{Corresponding author.}}
\author[a]{T.J. Roberts,\note{Also at Muons, Inc.}}
\author[a]{J.D. Phillips}
\author[b]{and R.D. Reasenberg}

\affiliation[a]{Illinois Institute of Technology,\\3101 S. Dearborn St., Chicago, IL 60616, USA}
\affiliation[b]{Center for Astrophysics and Space Sciences, University of California, San Diego,\\9500 Gilman Dr., La Jolla, CA 92093, USA}

\emailAdd{kaplan@iit.edu}

\abstract{We describe new results from the semiconductor-laser tracking frequency gauge, an instrument that can perform sub-picometer distance measurements and has applications in gravity research and in space-based astronomical instruments proposed for the study of light from extrasolar planets. Compared with previous results, we have improved incremental distance accuracy by a factor of two, to 0.9\,pm in 80\,s averaging time, and absolute distance accuracy by a factor of 20, to 0.17\,$\mu$m in 1000\,s. After an interruption of operation of a tracking frequency gauge used to control a distance, it is now possible, using a nonresonant measurement interferometer, to restore the distance to picometer accuracy by combining absolute and incremental distance measurements.}


\begin{document}
\maketitle
\flushbottom

\section{Introduction}
\label{sec:intro}

The semiconductor laser tracking frequency gauge (SL-TFG or, simply, TFG) was developed by a group at the Harvard--Smithsonian Center for Astrophysics (CfA) as a component of ultraprecise astronomical instruments and for tests of the equivalence principle~\cite{RSI,OL}. It has demonstrated unrivaled precision of both absolute and incremental (i.e., relative) position measurement. It employs Pound--Drever--Hall (PDH) locking of  an inexpensive IR communications laser to a null of an optical  interferometer, thereby converting position differences  among multiple TFGs into beat frequencies, which are conveniently in the RF band, facilitating their precise measurement. 
Previous tests carried out using a nonresonant interferometer at CfA~\cite{OL} have established precision of 2\,pm with 1\,s averaging time, and 0.14\,pm (improving to 46\,fm at 50\,s averaging time) using a Fabry--Perot  cavity as interferometer. 

Here we report new measurements using the same apparatus, after disassembly at CfA and reassembly and further development by new collaborators at Illinois Institute of Technology. These measurements show improved performance for both incremental distance (the change of distance since the start of operation) and absolute distance, which with sufficient accuracy allows resetting of the distance to the same interferometer fringe after an interruption. Incremental distance is measured to pm precision, while absolute distance is measured to $\sim$\,100\,nm. For the first time in a nonresonant measurement interferometer (MI), absolute distance is measured to an accuracy $\sigma(L)<\lambda/6$, the criterion for returning reliably to the same fringe after an interruption. Nonresonant measurement interferometers are much easier to work with than resonant ones, so are likely to be required in a space-borne astronomical instrument. The success of this effort emphasizes the simplicity and robustness of the instrument concept and implementation. The goals of the present effort include completing the ``G-POEM'' test~\cite{GPOEM1} of Lorentz violating matter--gravity couplings in the context of the Standard Model Extension~\cite{Kostelecky} and a new experiment (the Muonium Antimatter Gravity Experiment, MAGE) to measure the gravitational acceleration of unstable antimatter (a beam of slow muonium)~\cite{Kaplan,Atoms}, both requiring extreme precision of position measurement.

\section{Principle of the instrument}

\begin{figure}
\centerline{\includegraphics[width=6in,trim=120 120 30 160 mm,clip]{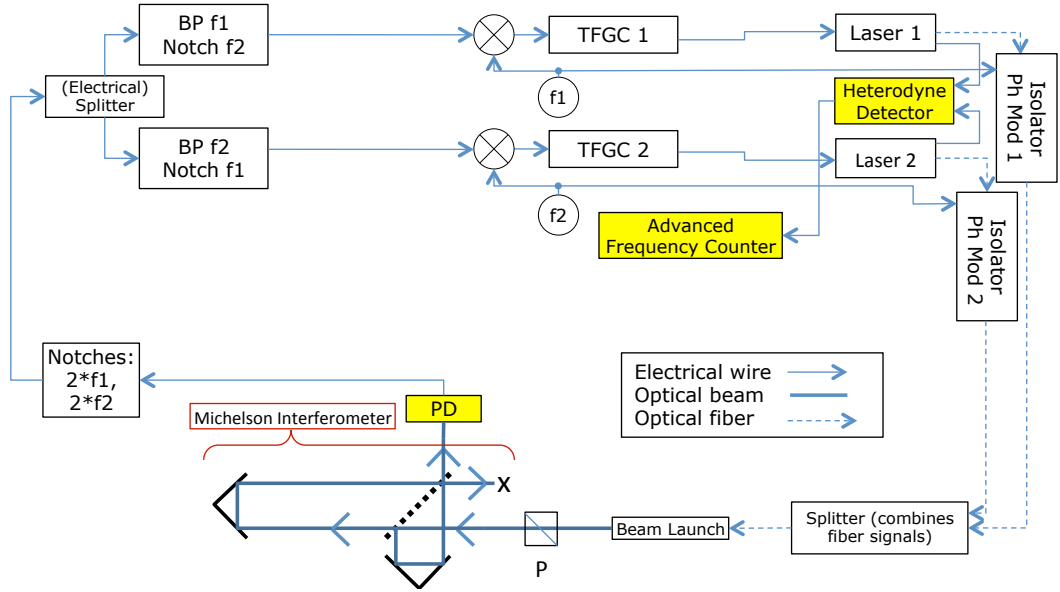}}
\caption{Block diagram of IIT 2-TFG test using Michelson interferometer. Function generators f1 and f2 (at upper-center of diagram) phase-modulate laser 1 and 2 signals and supply local-oscillator signals that are mixed with the electrical output of the photodetector in order to demodulate it at the two frequencies. Polarizer P selects the desired polarization. Notch filter at lower-left filters out  harmonics of f1 and f2, and bandpass/notch filters at upper-left and mixers operating at f1 and f2 separate the two electrical signals so that each TFG controller (TFGC) controls the correct laser. The heterodyne detector beats the two laser signals against each other, with beat frequency counted by the Advanced Frequency Counter, which is read out via ethernet for subsequent data analysis.}\label{fig:2-TFG-test}
\end{figure}

In a TFG, a single variable-frequency laser is PDH-locked to an interferometer null via a feedback system, which has a closed-loop bandwidth of $\sim$\,100\,kHz. In order to increase the available range of position measurement, as well as to provide a built-in absolute-length calibration, the feedback controllers (TFGCs)  are designed to allow the TFG to frequency-hop between adjacent interferometer nulls. $n$ distances can be measured with $n$ TFGs, plus one reference laser locked to a stable reference interferometer; the reference laser likely uses the same hardware as the TFGs. In our test setup, the two locked lasers are identical, so either could be viewed as the reference laser.  Our test employs two SL-TFGs (Fig.~\ref{fig:2-TFG-test}), each locked to a different null of the same  interferometer. We adjust the lasers to have wavelengths that are one or more fringes apart and count the resulting beat frequency when both beams impinge on the same photodetector. 

\subsection{Test details}
In practice, the two laser beams~\cite{lasers} are each frequency-modulated\footnote{For a sinusoidal modulating waveform (as is the case here), frequency and phase modulation are identical.} at a different (8.5 and 11.5\,MHz) radio frequency, then combined onto a single optical fiber and launched together into a Michelson interferometer. After traversing both optical paths of the interferometer, the combined beams  impinge together on a photodetector~\cite{PD}, and the resulting electrical signal is demodulated at the two modulation frequencies in order to separate the two signals for processing in their respective TFG controllers. The lasers operate at 1560\,nm, the path-length difference between the two Michelson arms is 0.58\,m, and the laser frequency difference when locked one fringe apart is 258\,MHz.

\begin{figure}
\centerline{\includegraphics[width=\textwidth]{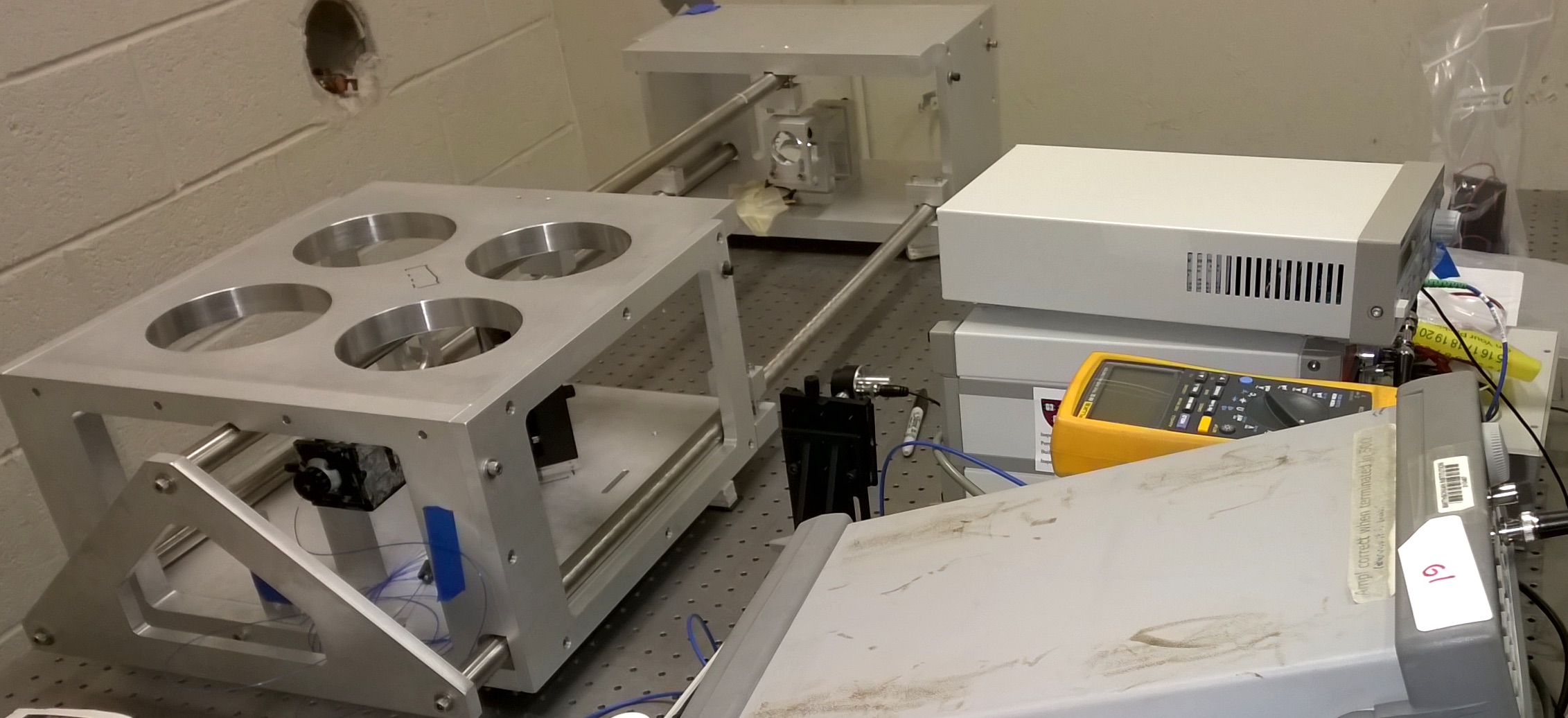}}
\caption{Photo of Michelson interferometer taken during initial setup at IIT; note long-arm corner cube visible towards top-center of photo, and beam launcher at lower-left with blue optical fiber entering it. For thermal stability, the interferometer assembly is supported on three Invar rods. The assembly sits on rubber pads on the optical table.}\label{fig:Michelson}
\end{figure}

\section{Results}
\subsection{Incremental distance measurement}

\begin{figure}
\centerline{\hspace{-.15in}\includegraphics[width=6in
]{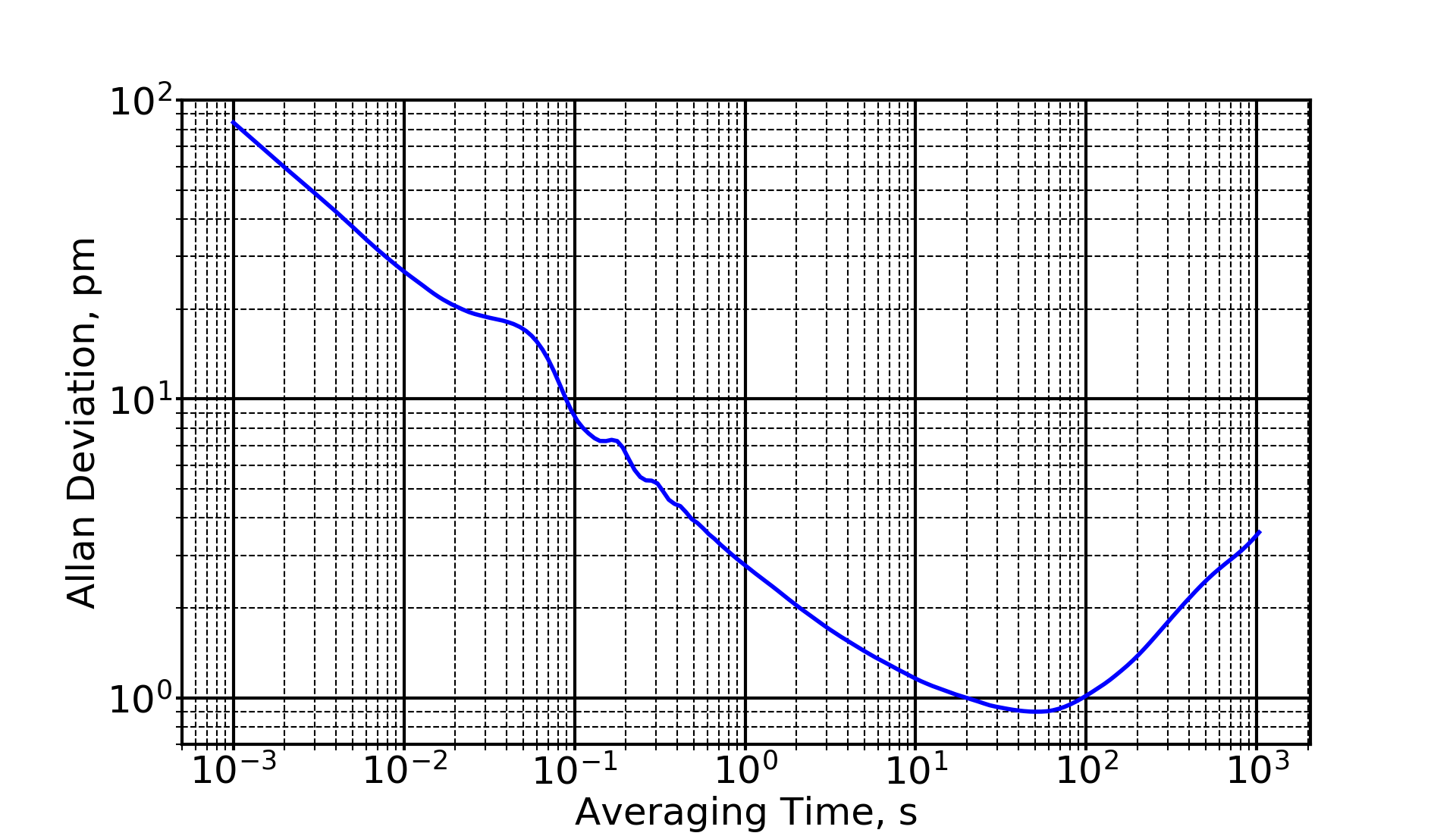}}
\caption{Allan deviation indicating TFG incremental-distance precision vs averaging time.}\label{fig:Allan}
\end{figure}

We have taken several records of advanced frequency counter (AFC) data in the configuration just described. Figure~\ref{fig:Allan} shows the Allan deviation derived from a recent AFC data record taken over 2 hours of continuous operation. The frequency change for a half-wavelength distance change is the free spectral range, $\Phi=c/2L$. For our MI, $\Phi=258$\,MHz. Changes $\delta f$ in the measured beat frequency are related to distance changes by 
\begin{equation}\label{eq0}
\frac{\delta f}{\delta L}=\frac{2\Phi}{\lambda}=\frac{331\,{\rm Hz}}{\rm pm}\,.
\end{equation}
Careful attention to electronic noise reduction, optical alignment, and focusing of the beams in the short and long Michelson arms has resulted in improved precision compared to that achieved at CfA by a factor $>2$, and we now consistently obtain incremental-distance precision of 1\,pm or less at $\sim\,$10$^2$\,s averaging time.

\begin{figure}
\centerline{\hspace{-.15in}\includegraphics[width=5.9in
]{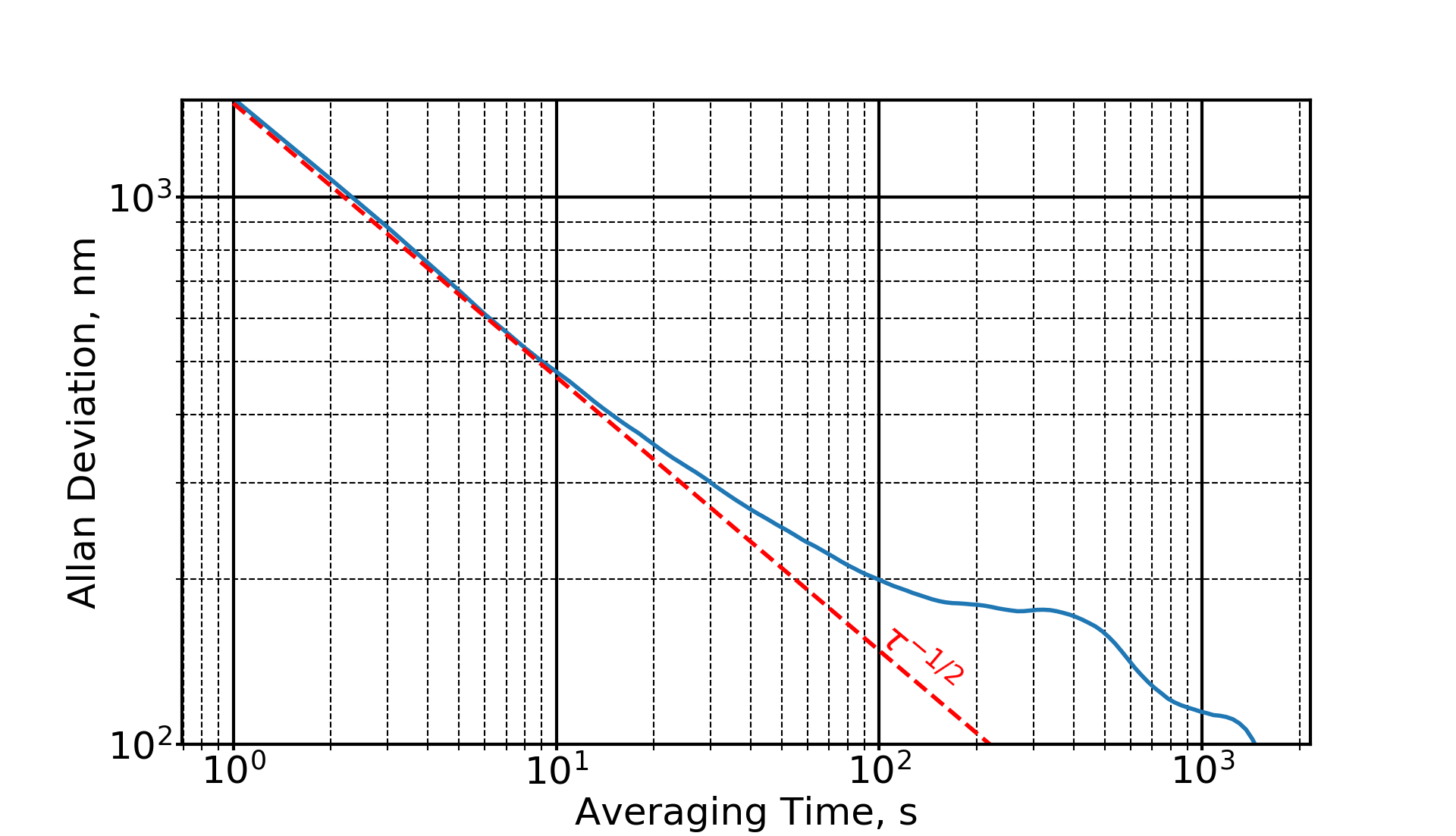}}
\caption{Allan deviation indicating TFG absolute-distance precision vs averaging time.}\label{fig:Allan-abs}
\end{figure}

\subsection{Absolute distance measurement}

The difference of frequencies of two lasers locked to fringes $k$ orders apart is
\begin{equation}\label{eq1}
\Delta \nu=\frac{kc}{2L}+\delta_1-\delta_2\,,
\end{equation}
where $c$ is the speed of light, $L$ is the interferometer arm-length difference, and $\delta_1$ and $\delta_2$ are the offsets of lasers 1 and 2 from the centers of their respective fringes. Eq.~\ref{eq1} provides the basis for determining absolute distance in terms of the speed of light and a measured radio frequency~\cite{RSI}. The offsets are due to optical and electronic effects as described in Sec.~\ref{sec:syst}. By using two or more values of $k$, we subtract the offsets to the extent that they are independent of time and of the MI order to which the laser is locked. We access multiple values of $k$ by hopping one laser between fringes at $\sim$\,1\,s intervals. 

In this paper, we present for the first time TFG measurements of absolute distance with $\Delta k>1$ and spanning more than two values of $k$. These improvements, as well as improvements in the accuracy of locking the lasers to MI fringes, result in the absolute distance measurements of Fig.~\ref{fig:Allan-abs}, which shows an Allan deviation of absolute distance having a minimum of 0.17\,$\mu$m at an averaging time of $\approx$\,250\,s. The Allan deviation shows little rise at longer averaging time, out to the longest time tested, $\approx$\,1000\,s. This accuracy is 0.3\,ppm of the arm length difference. At this wavelength, air contributes 158\,$\mu$m to the measured distance. The observed variation could be caused by a 0.3\,K temperature change, or a pressure change of 1\,mbar. Further improvements in absolute distance uncertainty will need to be tested in vacuum, in a helium atmosphere, or with temperature and pressure monitoring.

Consider the application of the TFG to a space-borne astronomical instrument for detection of light from an extrasolar planet, such as LUVOIR~\cite{LUVOIR} or HDST~\cite{HDST}. These missions require suppression of light from the host star by 10 orders of magnitude. This in turn may require structure held stable to picometer accuracy for periods of hours. During this time, starlight photons that leak into the ``dark hole'' are collected, in order to guide the adjustment of the starlight-suppressing optics. It is convenient if this stability is available across disturbances such as slews of pointing direction, or even interruptions such as power failures. 

If absolute distance is measured to an accuracy 
\begin{equation}\label{eq2}
\sigma(L)<\frac{\lambda}{6}\,,
\end{equation}
then with 3$\sigma$ confidence, after an interruption, operation can resume at the same interferometer fringe, meaning that distances can be reset to the full (pm-scale) accuracy of the incremental distance measurements. The absolute distance measurements presented here meet the criterion \ref{eq2}, for the first time for the TFG and as far as we know, for the first time ever in an instrument that is also capable of picometer incremental distance accuracy. 

Several effects cause the offsets $\delta_1$ and $\delta_2$ in Eq.~\ref{eq1}. Constant offsets can be tolerated, but variations mimic distance changes and must be eliminated to the extent possible. We have investigated the variability of offsets using measurements separated by several differing numbers of fringes. We find that the constant offsets can be of the order of 100--200\,kHz in our test setup. We infer the variation from the incremental distance measurements shown in Fig.~\ref{fig:Allan}, which shows a variation of 330\,Hz at an averaging time of 60\,s (see Eq.~\ref{eq0}). We discuss sources of this variation of offset below.

\section{Systematic Effects}\label{sec:syst}

In the PDH method, the variable-frequency laser carrier is purely frequency-modulated at a frequency $f_m$ before it impinges on the measurement interferometer (MI). When the carrier is tuned to the fringe minimum, the MI response is symmetrical. The signal emerging from the MI and impinging on the photodetector (PD in Fig.~\ref{fig:2-TFG-test}) is amplitude modulated only at $2f_m$. When the carrier is displaced from the fringe center, the signal emerging from the MI is amplitude modulated at $f_m$, and the component of AM at $f_m$ indicates the sign and magnitude of the detuning. To recover this component, each TFG has a bandpass/notch filter and RF mixer (Fig.~\ref{fig:2-TFG-test}), producing a detuning signal at baseband. The TFG Controller (TFGC) amplifies this signal and uses it to control the laser.

In addition to the FM-to-AM conversion by the MI, there is unwanted FM-to-AM conversion due to weak reflections of the laser signal on its way from the phase modulators to the photodetector. Pairs of such reflections create weak ``spurious etalons,'' which convert a small amount of the FM to AM. This AM adds to the desired signal, and contributes to the offsets. An offset in the measured signal of 1\,pm\,$\sim10^{-6}\lambda$ results from a spurious electric field of only $\sim10^{-5}$ of the carrier, or an optical intensity for the spurious round-trip light of $10^{-10}$ that of the desired light. Suppressing spurious reflections by this much requires care. 

The optical path length of these spurious etalons varies with temperature and even air density, varying their phase relative to the carrier. This causes the contributions to the offsets to vary. We believe that spurious AM is a major, and perhaps the dominant, contributor to the variation seen in Fig.~\ref{fig:Allan} and~\ref{fig:Allan-abs}. 

In addition to spurious AM, electrical errors introduced at the mixer outputs and in the TFGC input amplifiers can contribute to offsets. We have performed tests with the MI blocked, recording the TFGC output when it is configured as an amplifier. This test is sensitive to electrical offsets in all components including the mixer, operating with the same RF local oscillator as during distance measurements. By sweeping the laser frequency, it is also sensitive to offsets due to spurious AM. Tests have indicated little contribution from electronic errors, but clear contributions from spurious AM. For the data presented here, we reduced the spurious AM by carefully directing spurious reflections, e.g., those from the nominally non-reflecting side of the MI's beamsplitter, away from the desired beams. 

We have also investigated the time- and fringe-dependence of offset by hopping over a range of up to 12 fringes, dwelling at each fringe long enough to obtain a useful measure of offset. We will discuss these findings further in a subsequent publication.

To further improve the TFG accuracy, particularly at long times, we can work to further reduce spurious etalons. Alternatively, we can collect a sample of the light with a beamsplitter just before it enters the MI. We can detect the AM with a photodetector, filter, and mixer similar to those shown in Fig.~\ref{fig:2-TFG-test}. This signal can be used either to remove the spurious AM in post-processing, or to feed back to an amplitude modulator to null the spurious AM. We have started assembling such a system, but have not yet tested it. 

\section{Conclusion}
	We have described new results from the Tracking Frequency Gauge, which has applications in gravity research and in astronomical instruments for observing the light from extrasolar planets. Compared with our previous results, we have improved incremental distance accuracy by a factor of two and absolute distance accuracy by a factor of 20. After an interruption of operation of a TFG used to control a distance, it is now possible using a nonresonant interferometer to restore the distance to picometer accuracy by combining absolute and incremental distance measurements. We plan further refinement of error estimates and investigation of mechanisms in order to build a thorough error model, which will improve predictability of performance. 

\acknowledgments

We thank the Smithsonian Astrophysical Observatory (of the Harvard--Smithsonian Center for Astrophysics) for donation of  two TFG laser gauges, IIT undergraduate B. Carlson for help with TFG setup and data acquisition, as well as the Physics Dept., College of Science, BSMP~\cite{BSMP}, and IPRO~\cite{IPRO} programs at Illinois Institute of Technology for their support of our research.



\begin{thebibliography}{99}

\bibitem{RSI}
J.D. Phillips and R.D. Reasenberg, {\em Rev. Sci. Instrum.} {\bf 76} (2005) 
064501.

\bibitem{OL}
R. Thapa, J.D. Phillips, E. Rocco, and R.D. Reasenberg, {\em Opt. Lett.} {\bf 36} (2011) 3759.\\

\bibitem{GPOEM1}
  R.D. Reasenberg, J.D. Phillips, {\em A Laboratory Test of the Equivalence Principle as Prolog to a 
 Spaceborne Experiment, Proceedings From Quantum to Cosmos: Fundamental Physics Research in Space}, {\em IJMP} D {\bf 16}, \#12a, 2245 (2007); also in
 R.D. Reasenberg, J.D. Phillips, in {\bf From Quantum to Cosmos: Fundamental Physics Research in Space}, ed.\ Slava G. Turyshev, World Scientific, 2009. ISBN \#9789814261210, pp.\ 373--386.
  
\bibitem{Kostelecky}
V.A. Kosteleck\'y, J.D. Tasson, {\em Matter-gravity couplings and Lorentz violation}, {\em Phys.\ Rev.}\ D, {\bf 83}, 016013 (2011).

\bibitem{Kaplan}
D.M. Kaplan {\it et al.}, {\em Antimatter gravity with muonium}, prepared for Proc.\ 3rd International Workshop on Antimatter and Gravity (WAG2015), University College London, UK, 5--7 August 2015. arXiv:1601.07222 [physics.ins-det].

\bibitem{Atoms}
A. Atognini {\it et al.}, {\em Studying Antimatter Gravity with Muonium}, Atoms, \url{http://www.mdpi.com/journal/atoms}, to appear. 	arXiv:1802.01438 [physics.ins-det].

\bibitem{lasers}
DiCOS model RTL-2100, TeraXion, \url{www.teraxion.com}.

\bibitem{PD}
New Focus model 1611, Newport Corp., \url{www.newport.com}.

\bibitem{LUVOIR}
B.M. Peterson, D. Fischer [LUVOIR Science and Technology Definition Team], {\em The Large Ultraviolet/Optical/Infrared Surveyor (LUVOIR)}, American Astronomical Society, AAS Meeting \#229, 405.04 (2017).

\bibitem{HDST}
{\em From Cosmic Births to Living Earths}, Associated Universities for Research in Astronomy, 2015. Available as arXiv:1507.04779, \url{arxiv.org/abs/1507.04779}.

\bibitem{BSMP}
Brazilian Scientific Mobility Program, \url{https://www.iie.org/Programs/Brazil-Scientific-Mobility}.

\bibitem{IPRO}
IIT Inter-Professional Projects Program, \url{http://ipro.iit.edu}.




\end{thebibliography}
\end{document}